\documentclass{aa}
\usepackage{graphicx}
\usepackage{txfonts}
\usepackage{epsfig}
\usepackage{natbib}

\newcommand{\hd}{HD~81809}   
   
\newcommand{\caii}{Ca\,{\sc ii}}
\newcommand{\lx}{$L_{\rm X}$}  
\newcommand{\xmm}{XMM-\emph{Newton}}

\begin{document} 
\title{Fifteen years in the high-energy life of the solar-type star HD\,81809}
\subtitle{XMM-Newton observations of a stellar activity cycle}
\author{S.\ Orlando\inst{1} 
   \and F.\ Favata\inst{2} 
   \and G.\ Micela\inst{1} 
   \and S.\ Sciortino\inst{1}
   \and A.\ Maggio\inst{1}
   \and J.\,H.\,M.\,M. Schmitt\inst{3} 
   \and J.\ Robrade\inst{3}
   \and M.\ Mittag\inst{3}
}

\offprints{S. Orlando}

\institute{INAF -- Osservatorio Astronomico di Palermo, Piazza del Parlamento 1, I-90134 Palermo, Italy\\
\email{orlando@astropa.inaf.it}
\and European Space Agency, 8-10 rue Mario Nikis, 75738 Paris cedex 15, France
\and Universit\"at Hamburg, Hamburger Sternwarte, Gojenbergsweg 112, 21029 Hamburg, Germany
  }
\date{Received date / Accepted date}

\abstract
%Context
{The modulation of the activity level of solar-like stars is commonly
revealed by cyclic variations in their chromospheric indicators,
such as the {\caii} H\&K ``$S$-index'', similarly to what is observed
in our Sun. However, while the variation of solar activity is
reflected also in the cyclical modulation of its coronal X-ray
emission, a similar behavior has only been discovered in a few stars
other than the Sun.}
%Aims
{The data set of the long-term \xmm\ monitoring program of \hd\ is
analyzed to study its X-ray cycle, to investigate if the latter is
related to the chromospheric one, to infer the structure of the
corona of \hd, and to explore if the coronal activity of \hd\ can
be ascribed to phenomena similar to the solar ones and, therefore,
considered an extension of the solar case.}
%Methods
{We analyze the observations of \hd\ performed with \xmm\ with a
regular cadence of 6 months from 2001 to 2016 and representing 
one of the longest available observational baseline ($\sim
15$~yr) for a solar-like star with a well-studied chromospheric
cycle (with a period of $\sim 8$~yr). We investigate the modulation
of coronal luminosity and temperature and its relation with the
chromospheric cycle. We interpret the data in terms of a mixture
of solar-like coronal regions, adopting a methodology originally
proposed to study the Sun as an X-ray star.}
%Results
{The observations show a well-defined regular cyclic modulation of
the X-ray luminosity that reflects the activity level of \hd.
The data covers approximately two cycles of coronal activity; the
modulation has an amplitude of a factor of $\sim 5$ (excluding
evident flares, as in the June 2002 observation) and a period of
$7.3\pm 1.5$~yr, consistent with that of the chromospheric cycle. We
demonstrate that the corona of \hd\ can be interpreted as an extension
of the solar case and it can be modeled with a mixture of solar-like
coronal regions along the whole cycle. The activity level is mainly
determined by a varying coverage of very bright active regions,
similar to cores of active regions observed in the Sun. Evidence
of unresolved significant flaring activity is present especially in
proximity of cycle maxima.}
%Conclusions
{}


\titlerunning{The X-ray cycle in the solar-type star  HD\,81809}
\authorrunning{S. Orlando et~al.}

\maketitle

\section{Introduction}
\label{sec:intro}

Cyclic modulations in the activity level of late-type stars are
commonly detected through chromospheric indicators, as the {\caii}
H\&K activity indicator (the ``$S$-index''). The long-term monitoring
of the {\caii} $S$-index measured in a large sample of stars with
the Mt. Wilson 100-inch telescope has shown that solar-like cycles
(with periods in the range $2.5 - 25$~yr) are present in many ($\sim
60$~\%) main-sequence stars with spectral types between F and M
(\citealt{1995ApJ...438..269B}).  Most often the cycles are detected
in intermediate-activity stars, whereas less active stars show no
evident variability, suggesting that they are in a ``Maunder
minimum''-like state (i.e. very small long-term variability), and
the most active stars are dominated by irregular variability rather
than by cyclic modulation. The Sun fits well in this sample with
the well known 11-yr cycle, evident in the {\caii} $S$-index
(\citealt{1992PASP..104.1139W}).

In the Sun, the changes in its activity level are also evident
through a cyclic variation in the coronal emission: the solar X-ray
luminosity follows a cycle with the same periodicity as the
chromospheric one and an amplitude between one and two orders of
magnitude in the [$0.1 - 3$]~keV band (e.g. \citealt{opr01,
2003ApJ...593..534J}). It is worth to mention that solar cycles are
similar to each other but not identical, neither in duration, nor
in magnitude, and long-term changes do occur. In late-type stars
other than the Sun, cyclical variations in the coronal emission
have been found in recent years and only for few cases. The main
reason is that typical targets of X-ray observations have been stars
with a high activity level which, however, show modest (if any)
amounts of long-term variability with no evidence of cyclic modulation
(e.g.  \citealt{1998ASPC..154..223S, 2004A&A...417..651S}). A first
hint that cyclic chromospheric activity implies cyclic coronal
activity was found - on the base of statistical arguments - in
low-activity stars by using the ROSAT all-sky survey (RASS)
(\citealt{1996A&A...305..284H}). To date, clear cyclic variations
in the coronal emission have been detected in three binary stars:
61 Cyg A (\citealt{2003A&A...406L..39H, 2006A&A...460..261H}),
$\alpha$ Cen (\citealt{2005A&A...442..315R, 2012A&A...543A..84R,
2017MNRAS.464.3281W}), and \hd\ (\citealt{2004A&A...418L..13F} -
Paper I - \citealt{2008A&A...490.1121F} - Paper II). More recently
a cyclic modulation in X-rays which follows the contemporaneous
chromospheric cycle was also discovered in the young ($\approx
600$~Myr) late-type star $\iota$ Hor (\citealt{2013A&A...553L...6S}).

These stars provide the best available data set to investigate the
cyclical variations in the coronal emission of late-type stars.
One of the longest available observational baseline ($\sim
15$~yr) is that for \hd\ which is also the system showing the
clearest evidence of a cyclical modulation in its coronal emission
(Paper II). \hd, at a nominal distance of $\approx 31.2$~pc, is
known to be a binary system with an orbital period of $\sim 35$~yr
and a maximum separation of $\sim 0.4$~arcsec
(\citealt{2000A&AS..145..215P, 2015AJ....150...50T}; for more details
see Paper I). The two components have spectral types G2 and G9, and
apparent magnitudes $V_1 = 5.8$ and $V_2 = 6.8$, respectively; their
masses are $M_1 = 1.7 \pm 0.64$\,M$_{\odot}$ and $M_2 = 1.0 \pm
0.25$\,M$_{\odot}$, respectively. Both components are slow rotators,
with $v \sin i \approx 3$~km~s$^{-1}$ (\citealt{1982ApJ...263..239S}).
\hd\ is generally considered to have an age close to that of the
Sun. However, \cite{2004A&A...418L..13F} have suggested that the
stars are not on the main sequence, but rather subgiants.  Although
\hd\ is binary in nature, this does not hamper studies on the cyclic
modulation of coronal emission: the large physical separation of
the two stars makes tidal effects negligible, and the low rotational
velocity confirms that this is a non-interacting system. The very
clear cyclical behavior of \hd\ in the {\caii} $S$-index (as evident
from the Mt. Wilson program data; \citealt{1995ApJ...438..269B})
and in X-rays (see Paper II) points to the activity being dominated
by one of the two components.

The chromospheric cycle of \hd\ has a regular modulation similar
to that of the Sun but with a shorter period of 8.2~yr. The analysis
of the {\xmm} data from 2001 to 2007 (thus covering six years of
the cycle) discussed in Paper II has shown that a well-defined cycle
seems to be present also in X-rays and appears to follow that in
the {\caii} $S$-index. Our previous analysis did not allow us to
constrain accurately the amplitude of the X-ray cycle because the
only maximum recorded was not well sampled and one of the data at
the maximum was most likely affected by a flaring event which led
to an overestimation of the flux. We derived an average luminosity
approximately one order of magnitude higher than in the Sun. The
previous analysis covered most (but not all) of one cycle describing
a maximum in 2002 and a minimum in 2005. However, no evidence of a
second maximum was present in the data and no indication that the
next cycle has similar duration and amplitude as the first one.
Here we analyze all available {\xmm} observations, from 2001
to 2016, thus covering almost two coronal cycles. We compare the
two cycles exploring possible significant changes (if any) and
interpret the data in terms of a mixture of coronal regions observed
in the Sun, adopting a methodology extensively applied to study the
``Sun as a star'' (\citealt{opr00, porrh00}).  In Sect.~\ref{sec2},
we present the {\xmm} data; in Sect.~\ref{sec3} we describe the
data analysis and discuss the results; and finally in Sect.~\ref{sec4}
we draw our conclusions.

\section{Observations}
\label{sec2}

\hd\ has been selected as target to investigate cyclic variations
in the coronal emission because it has: (i) enough X-ray flux at
Earth to allow an accurate analysis of the data, and (ii) a well
defined cycle in the {\caii} $S$-index (from Mt. Wilson program
data; \citealt{1995ApJ...438..269B}) with a period of 8.2~yr which
allows to study few cycles in a reasonable time. The strategy to
sample the cyclic behavior of \hd\ was based on the expected cycle
duration of 8.2~yr and considered two observations per year, spaced
by 6 months (see Table~\ref{tab1}). Since April 2001, the observations
have been executed regularly each year in the Spring (between April
and June) and in the Fall (between September and November), with
the only exception of the Spring 2006 season, due to a technical
problem that prevented the observation from being carried out. The
observations have different duration (ranging between 6 and 30~ks;
see Table~\ref{tab1}), because the exposure time has been adapted
along the program to follow the variation in luminosity of the
target. All observations up to May 2003 were taken with the medium
filter; all the others are taken with the thick filter (this is due
to a policy change of the observatory).

\section{Analysis} 
\label{sec3}

All data sets are processed homogeneously using the SAS package.
As a first step, we remove high background time intervals from the
observations. The data collected on April 2001 is rejected from the
analysis because the observation is affected by a very high background
level (much higher than in the other observations) which, coupled
with the short duration of the observation, resulted in a high noise
level which could not be filtered out. Then, from the processed
data, we extract lightcurves and spectra. We estimate the parameters
characterizing the spectra (i.e. effective temperature and emission
measure) by fitting simultaneously pn and MOS spectra, using XSPEC,
with thermal plasma radiative loss models. To this end, we adopt
Astrophysical Plasma Emission Code (APEC) models, considering the
metal abundance frozen at $Z = 0.3\, Z_{\odot}$, as in Papers I and
II. As discussed in Paper II, the interstellar absorption is
consistent with a zero value, and we do not include it in the fit.

\subsection{Characterizing the coronal plasma using model parameter
estimation}
\label{sec3.1}

The spectra are first matched to an isothermal model, following
the approach of Paper II. This allows a consistent comparison of
our results with those of the previous papers (Papers I and II). We
find that the results of the fit during the two cycles are similar
to those found in Paper II, with a modest range of variability in
best-fit temperature and a good correlation between temperature and
stellar X-ray luminosity. Most of the spectra are well fit
with a single-temperature model; however, in some cases, the $\chi^2$
of the fit yields a very low probability as, for instance, for the
June 2002 data set when, most likely, a bright flare was present
(see Paper II for more details). In the light of this, we explore
the possibility to fit the spectra with more complex models that
included two temperatures plasma components. In the fit, the
temperatures and emission measures are allowed to vary and all the
spectra are well fit (including the June 2002 spectrum). Table~\ref{tab1}
reports the spectral parameters resulting from the fit for the
complete observation timeline from November 2001 to October 2016.
The Table reports also the emission measure-weighted temperature,
$T_{ave}$, defined as

\begin{equation}
T_{ave} = \frac{{\rm EM}_1 T_1 + {\rm EM}_2 T_2}{{\rm EM}_1+{\rm EM}_2}
\label{tave}
\end{equation}

\noindent
where $T_1$ and $T_2$ are the temperatures of the two isothermal
components and EM$_1$ and EM$_2$ are their emission measures.

\begin{figure}
  \begin{center}
    \leavevmode
        \epsfig{file=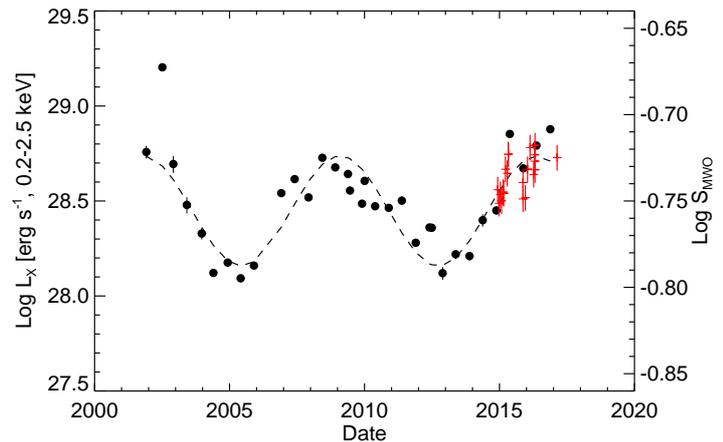, width=9.cm}
	\caption{Evolution of the X-ray luminosity in the $[0.2-2.5]$
	keV band of \hd\ from November 2001 to October 2016, covering
	almost two cycles of coronal activity. S$_{MWO}$ for the
	chromospheric \caii\ activity is superimposed (red crosses,
	right hand scale), derived from data collected with the
	TIGRE telescope.  In most of the X-ray data points the
	uncertainty on the X-ray luminosity is smaller than the
	symbol size; the uncertainty on the S-index is shown. The
	dashed curve is a log-sinusoidal function (with a period
	of $7.3$~yr) which fits the X-ray data (see the text).}
  \label{fig1}
\end{center} \end{figure}

For all spectra, we compute the X-ray luminosity in the $[0.2-2.5]$
keV band. Figure~\ref{fig1} shows the light curve of the X-ray
luminosity of \hd\ together with a log-sinusoidal fit to the data
(dashed line in the figure). On average the observations show a
long-term variability in good agreement with the cyclical behavior
of \hd\ in the {\caii} $S$-index (see \citealt{1995ApJ...438..269B}
and Paper II). The coherent behavior of this variability is consistent
with a regular cyclic modulation in the coronal activity level. Our
data cover approximately two cycles of coronal activity. A first
maximum occurred approximately in 2001, although a significant level
of uncertainty is present given that this maximum has not been
properly sampled by our observations (see also Paper II). Then the
luminosity decreased to a minimum around 2005 and increased toward
a second broad maximum that occurred in 2009-2010.  The X-ray
luminosity in the second maximum is very close to the values observed
in late 2001, supporting the assumption that the first observed
maximum was in 2001. Around 2013 the luminosity reached a second
minimum with X-ray luminosity similar to that of 2005 and, then,
increased again toward a new maximum in 2016. From the fitting of
the data with a sinusoidal function, we find that the cyclic
modulation of the X-ray luminosity has an amplitude of a factor of
$\sim 5$, if one ignores the apparently anomalous June 2002
observation, and a period of $7.3\pm 1.5$~yr very close to that of
the chromospheric cycle. According to these estimates, the star is
currently at its maximum. The analysis shows that the two cycles
appear very similar each other and suggest a regular coronal cycle
in \hd\ analogous to that observed in the Sun. We note that the
amplitude of the cycle (about a factor of 5) is much lower than
that of the Sun (which is about a factor of 50; \citealt{opr01}),
probably due to the higher level of activity of \hd. We
conjecture that this might reflect a dynamo more productive of
magnetic fields, so that a significant coronal activity (in terms
of presence of active regions, cores, and possibly flares) is always
present on the stellar surface (even during the minimum) at variance
with the solar case. This idea is supported by the evidence that
the X-ray flux of \hd\ at minimum is much larger (by a factor of
$\approx 6$) than the X-ray flux of the Sun at minimum (see also
Fig.~3 in Paper II), suggesting a significant coronal activity in
\hd\ even at minimum.

The June 2002 data point is the only most prominent exception to
the regular cyclic modulation in the coronal activity level with a
luminosity $3\times$ higher than during the maximum; this was most
likely due to a bright flare dominating the X-ray emission of \hd\
during the observation. In fact, the spectrum in June 2002 is
characterized by a second temperature component at $\sim 15.4$~MK
and corresponding emission measure comparable to that of the
colder component (see Table~\ref{tab1}). As discussed in detail in
Papers I and II, these characteristics point at a period of enhanced
flare-like activity, although the light curve extracted from this
observation (which however only spans approximately two hours) shows
no evidence of flare-like variability. The comparison of the June
2002 spectrum with the spectrum of the flaring Sun observed as a
star supports this interpretation (see Paper II and Sect.~\ref{sec3.2}
for more details). Other observations (e.g. April 2011 and April
2015) show similar high temperature components although with
corresponding emission measures lower than in June 2002.

\subsection{Chromospheric \caii\ activity measured by TIGRE}

In correspondence of the last maximum in the coronal activity level
of \hd, chromospheric \caii\ activity has been measured continuously
between 2014 and 2017 by TIGRE, a fully automatic 1.2~m telescope
located at the La Luz Observatory in central Mexico. TIGRE's main
instrument is the fibre-fed \'Echelle spectrograph HEROS with a
spectral resolution of R$\approx$20,000 and a spectral range from
$\approx 3800$~\AA~to 8800~\AA\ with a small gap of $\approx$100~\AA\
centered around 5800~\AA; a detailed description of the TIGRE
facility is given by \cite{2014AN....335..787S}.

The TIGRE facility is primarily designed to execute spectroscopic
long-term monitoring programs and one of these programs is the
long-term monitoring of active stars. The most commonly used activity
indicator is the Mount Wilson S-index (\citealt{1995ApJ...438..269B}).
With the TIGRE spectra we first derive an instrumental S-index
(S$_{\rm{TIGRE}}$) with an automatic procedure (see
\citealt{2016A&A...591A..89M} for a detailed description), which
works as follows: first, the radial velocity (RV) shift of the
\caii\ region is estimated via a cross correlation with a synthetic
spectrum and then a RV correction of the \caii\ region is performed.
Afterwards, the counts in the four band passes of the S$_{\rm{TIGRE}}$
index are integrated and the ratio between the counts of both line
centers and the quasi reference continua are calculated. This ratio
yields the S$_{\rm{TIGRE}}$-value, which is transformed into the
Mount Wilson scale with a linear relation derived by
\cite{2016A&A...591A..89M} through

\begin{eqnarray}
S_{\rm{MWO}} & = &
(0.0360\pm0.0029)+(20.02\pm0.42) S_{\rm{TIGRE}}. 
\label{trans_eq_hrt_s_mw_index}
\end{eqnarray}

\noindent
In this fashion our TIGRE S-indices can be readily compared to the
vast amount of literature available on the activity derived from
the \caii\ line cores. The changes in $S_{\rm{MWO}}$, measured
between 2014 and 2017, are shown in Fig.~\ref{fig1} (red crosses
superimposed to the X-ray lightcurve). We find that, in the period
analyzed, the chromospheric \caii\ activity level increases following
the increase in the coronal activity level. Unfortunately the TIGRE
data do not cover a full cycle of activity, so that we cannot derive
the period and amplitude of the cyclic modulation of chromospheric
\caii\ activity. In the next future, therefore, it is highly desirable
to continue the monitoring of \hd\ with both \xmm\ and TIGRE
instruments to enable a more precise comparison between coronal and
chromospheric activity variations.

\begin{table*}
  \begin{center}
    \caption{Best-fit spectral model parameters for the 33 \xmm\ observations
      of \hd, spanning 15 years.} 
    \begin{tabular}{cccccccc}
\hline\hline
Obs. ID & Obs. start UTC & Obs. duration & \lx\ & $T_1$ & $T_2$ & EM$_2$ / EM$_1$ & $T_{ave}$ \\
     & & [s] & [$10^{28}$~erg~s$^{-1}$] & [MK] & [MK] & &  [MK] \\\hline
\vspace{-0.25cm} \\
0032340301 &    2001-11-01 23:16:52  &   6402 &  $6.63^{+0.45}_{-0.45}$ & $3.69^{+0.38}_{-0.38}$ & $8.60^{+0.90}_{-0.90}$ & $0.54^{+0.20}_{-0.20}$ & $5.41^{+1.50}_{-1.50}$ \\
\vspace{-0.25cm} \\
0032341301 &    2002-11-02 18:04:22  &   5453 &  $18.4^{+0.30}_{-0.30}$ & $4.05^{+0.29}_{-0.29}$ & $15.4^{+0.34}_{-0.34}$ & $1.38^{+0.13}_{-0.13}$ & $10.6^{+0.66}_{-0.66}$ \\
\vspace{-0.25cm} \\
0032342201 &    2002-06-06 03:07:46  &   6395 &  $5.72^{+0.50}_{-0.50}$ & $2.83^{+1.21}_{-1.21}$ & $7.10^{+1.85}_{-1.85}$ & $0.60^{+0.60}_{-0.60}$ & $4.44^{+3.35}_{-3.35}$ \\
\vspace{-0.25cm} \\
0032342301 &    2003-05-03 11:10:15  &   8452 &  $3.49^{+0.29}_{-0.29}$ & $3.71^{+0.39}_{-0.39}$ & $8.20^{+1.34}_{-1.34}$ & $0.32^{+0.18}_{-0.18}$ & $4.80^{+1.65}_{-1.65}$ \\
\vspace{-0.25cm} \\
0143630701 &    2003-11-22 05:33:21  &   7208 &  $2.47^{+0.15}_{-0.15}$ & $2.76^{+0.50}_{-0.50}$ & $6.10^{+2.78}_{-2.78}$ & $0.35^{+0.39}_{-0.39}$ & $3.62^{+2.45}_{-2.45}$ \\
\vspace{-0.25cm} \\
0143630801 &    2004-04-29 22:09:49  &  22319 &  $1.53^{+0.06}_{-0.06}$ & $2.28^{+0.40}_{-0.40}$ & $4.77^{+1.10}_{-1.10}$ & $0.83^{+0.51}_{-0.51}$ & $3.41^{+1.65}_{-1.65}$ \\
\vspace{-0.25cm} \\
0202610801 &    2004-11-09 19:16:08  &  30708 &  $1.73^{+0.06}_{-0.06}$ & $3.68^{+0.19}_{-0.19}$ & $8.80^{+1.94}_{-1.94}$ & $0.13^{+0.09}_{-0.09}$ & $4.28^{+0.96}_{-0.96}$ \\
\vspace{-0.25cm} \\
0202611001 &    2005-05-03 19:19:09  &  28917 &  $1.43^{+0.06}_{-0.06}$ & $3.17^{+0.37}_{-0.37}$ & $6.34^{+1.34}_{-1.34}$ & $0.22^{+0.20}_{-0.20}$ & $3.75^{+1.50}_{-1.50}$ \\
\vspace{-0.25cm} \\
0202611601 &    2005-10-30 05:57:32  &  30119 &  $1.67^{+0.06}_{-0.06}$ & $3.67^{+0.25}_{-0.25}$ & $7.98^{+1.31}_{-1.31}$ & $0.28^{+0.15}_{-0.15}$ & $4.62^{+1.37}_{-1.37}$ \\
\vspace{-0.25cm} \\
0401880601 &    2006-11-05 23:41:06  &  22213 &  $4.02^{+0.11}_{-0.11}$ & $3.59^{+0.26}_{-0.26}$ & $7.12^{+0.72}_{-0.72}$ & $0.37^{+0.17}_{-0.17}$ & $4.55^{+1.30}_{-1.30}$ \\
\vspace{-0.25cm} \\
0401880701 &    2007-05-04 12:22:37  &  23209 &  $4.77^{+0.14}_{-0.14}$ & $3.82^{+0.14}_{-0.14}$ & $8.81^{+0.55}_{-0.55}$ & $0.41^{+0.08}_{-0.08}$ & $5.26^{+0.69}_{-0.69}$ \\
\vspace{-0.25cm} \\
0501930401 &    2007-11-09 16:42:54  &  16918 &  $3.82^{+0.10}_{-0.10}$ & $3.48^{+0.31}_{-0.31}$ & $7.25^{+1.15}_{-1.15}$ & $0.27^{+0.18}_{-0.18}$ & $4.27^{+1.51}_{-1.51}$ \\
\vspace{-0.25cm} \\
0501930501 &    2008-05-14 23:38:08  &  16917 &  $6.18^{+0.16}_{-0.16}$ & $3.44^{+0.18}_{-0.18}$ & $8.22^{+0.32}_{-0.32}$ & $0.59^{+0.09}_{-0.09}$ & $5.22^{+0.61}_{-0.61}$ \\
\vspace{-0.25cm} \\
0550061101 &    2008-11-06 18:13:31  &  16917 &  $5.50^{+0.15}_{-0.15}$ & $2.59^{+0.39}_{-0.39}$ & $5.40^{+0.34}_{-0.34}$ & $1.59^{+0.50}_{-0.50}$ & $4.32^{+0.93}_{-0.93}$ \\
\vspace{-0.25cm} \\
0550061201 &    2009-04-27 06:36:24  &  16915 &  $5.07^{+0.12}_{-0.12}$ & $2.84^{+0.29}_{-0.29}$ & $5.40^{+0.42}_{-0.42}$ & $1.07^{+0.33}_{-0.33}$ & $4.16^{+0.97}_{-0.97}$ \\
\vspace{-0.25cm} \\
0600820401 &    2009-05-24 23:00:31  &  11914 &  $4.15^{+0.16}_{-0.16}$ & $3.39^{+0.86}_{-0.86}$ & $6.19^{+2.99}_{-2.99}$ & $0.45^{+0.88}_{-0.88}$ & $4.26^{+5.33}_{-5.33}$ \\
\vspace{-0.25cm} \\
0600820501 &    2009-11-04 16:32:25  &  11919 &  $3.54^{+0.12}_{-0.12}$ & $2.97^{+0.23}_{-0.23}$ & $6.71^{+0.54}_{-0.54}$ & $0.41^{+0.11}_{-0.11}$ & $4.07^{+0.75}_{-0.75}$ \\
\vspace{-0.25cm} \\
0600820601 &    2009-12-08 07:22:42  &  13417 &  $4.67^{+0.16}_{-0.16}$ & $3.65^{+0.16}_{-0.16}$ & $8.13^{+0.71}_{-0.71}$ & $0.25^{+0.07}_{-0.07}$ & $4.54^{+0.71}_{-0.71}$ \\
\vspace{-0.25cm} \\
0600820701 &    2010-04-30 23:45:12  &  11915 &  $3.43^{+0.12}_{-0.12}$ & $3.52^{+0.19}_{-0.19}$ & $9.34^{+2.20}_{-2.20}$ & $0.12^{+0.08}_{-0.08}$ & $4.13^{+0.90}_{-0.90}$ \\
\vspace{-0.25cm} \\
0654550401 &    2010-10-31 17:48:38  &  13914 &  $3.37^{+0.13}_{-0.13}$ & $3.23^{+0.56}_{-0.56}$ & $6.46^{+0.51}_{-0.51}$ & $0.61^{+0.35}_{-0.35}$ & $4.45^{+1.82}_{-1.82}$ \\
\vspace{-0.25cm} \\
0654550501 &    2011-04-27 09:52:48  &  10921 &  $3.67^{+0.16}_{-0.16}$ & $3.84^{+0.13}_{-0.13}$ & $11.7^{+1.16}_{-1.16}$ & $0.19^{+0.04}_{-0.04}$ & $5.09^{+0.55}_{-0.55}$ \\
\vspace{-0.25cm} \\
0670320101 &    2011-11-02 11:38:16  &  12917 &  $2.20^{+0.09}_{-0.09}$ & $2.94^{+0.37}_{-0.37}$ & $6.22^{+2.35}_{-2.35}$ & $0.37^{+0.35}_{-0.35}$ & $3.82^{+2.16}_{-2.16}$ \\
\vspace{-0.25cm} \\
0670320601 &    2012-05-07 12:21:03  &  12918 &  $2.65^{+0.10}_{-0.10}$ & $2.93^{+0.27}_{-0.27}$ & $7.34^{+0.87}_{-0.87}$ & $0.35^{+0.12}_{-0.12}$ & $4.08^{+0.92}_{-0.92}$ \\
\vspace{-0.25cm} \\
0690800401 &    2012-06-07 00:38:40  &  14414 &  $2.64^{+0.11}_{-0.11}$ & $2.17^{+0.92}_{-0.92}$ & $4.35^{+0.78}_{-0.78}$ & $2.06^{+1.65}_{-1.65}$ & $3.63^{+2.05}_{-2.05}$ \\
\vspace{-0.25cm} \\
0690800501 &    2012-11-01 10:47:39  &  12017 &  $1.53^{+0.10}_{-0.10}$ & $3.19^{+1.60}_{-1.60}$ & $6.21^{+23.4}_{-23.4}$ & $0.10^{+1.33}_{-1.33}$ & $3.46^{+10.7}_{-10.7}$ \\
\vspace{-0.25cm} \\
0690800701 &    2013-04-27 23:47:21  &  17418 &  $1.92^{+0.09}_{-0.09}$ & $2.91^{+0.37}_{-0.37}$ & $6.89^{+0.51}_{-0.51}$ & $0.70^{+0.20}_{-0.20}$ & $4.55^{+1.01}_{-1.01}$ \\
\vspace{-0.25cm} \\
0720580801 &    2013-10-30 06:04:59  &  14600 &  $1.88^{+0.09}_{-0.09}$ & $2.52^{+0.38}_{-0.38}$ & $7.16^{+0.69}_{-0.69}$ & $0.51^{+0.14}_{-0.14}$ & $4.09^{+0.86}_{-0.86}$ \\
\vspace{-0.25cm} \\
0720580901 &    2014-04-28 11:30:24  &  26000 &  $2.90^{+0.19}_{-0.19}$ & $3.18^{+0.61}_{-0.61}$ & $6.22^{+0.70}_{-0.70}$ & $0.64^{+0.36}_{-0.36}$ & $4.36^{+1.78}_{-1.78}$ \\
\vspace{-0.25cm} \\
0741700401 &    2014-10-29 15:35:25  &  15999 &  $3.26^{+0.13}_{-0.13}$ & $1.76^{+0.40}_{-0.40}$ & $5.42^{+0.36}_{-0.36}$ & $1.61^{+0.57}_{-0.57}$ & $4.02^{+0.75}_{-0.75}$ \\
\vspace{-0.25cm} \\
0741700501 &    2015-04-30 11:36:57  &  22700 &  $8.24^{+0.22}_{-0.22}$ & $4.51^{+0.16}_{-0.16}$ & $11.2^{+0.62}_{-0.62}$ & $0.25^{+0.03}_{-0.03}$ & $5.84^{+0.44}_{-0.44}$ \\
\vspace{-0.25cm} \\
0760290401 &    2015-10-29 21:10:29  &  15000 &  $5.43^{+0.18}_{-0.18}$ & $3.43^{+0.43}_{-0.43}$ & $6.56^{+0.42}_{-0.42}$ & $0.67^{+0.30}_{-0.30}$ & $4.68^{+1.54}_{-1.54}$ \\
\vspace{-0.25cm} \\
0760290501 &    2016-04-25 07:18:57  &  18500 &  $7.16^{+0.20}_{-0.20}$ & $2.30^{+0.28}_{-0.28}$ & $6.01^{+0.34}_{-0.34}$ & $1.95^{+0.34}_{-0.34}$ & $4.75^{+0.61}_{-0.61}$ \\
\vspace{-0.25cm} \\
0783340401 &    2016-10-29 12:03:04  &  11000 &  $8.73^{+0.30}_{-0.30}$ & $2.16^{+0.26}_{-0.26}$ & $6.47^{+0.17}_{-0.17}$ & $1.68^{+0.21}_{-0.21}$ & $4.85^{+0.38}_{-0.38}$ \\
\vspace{-0.25cm} \\
\hline
    \end{tabular}
    \label{tab1}
  \end{center}
\end{table*}

\subsection{Characterizing the coronal plasma using a mixture of
solar-like coronal regions}
\label{sec3.2}

Although \hd\ cannot be considered a solar twin (because it is a
binary system and one of the two components has mass and radius
different from those of the Sun), we demonstrated in Paper II that
the observations can be interpreted in terms of a mixture of
solar-like active regions. At the base of this interpretation there
was the assumption that the X-ray activity of the system is dominated
by the more massive, larger companion (the primary component) with
a mass $M_1 \approx 1.7\,M_{\odot}$ and radius $R_1\approx 2-3\,R_{\odot}$.
This assumption is justified by the evident coherent, cyclic
variability observed (due to a dominant component in the system)
and by the impossibility to produce the observed levels of activity
with the smaller secondary component and filling factors smaller
than 1 (see also the discussion below on the comparison between the
X-ray luminosity of the Sun and that of \hd). Here we adopt the
same assumption and investigate if and how a mixture of solar-like
coronal regions can describe the corona of the primary component
of \hd\ along its cycle of activity.

In Paper II we adopted a methodology originally developed in the
context of the study of the ``Sun as a star'' (see \citealt{opr00,
porrh00}) and investigated if the cycle of \hd\ can be explained
in terms of solar-like structures by modeling the stellar corona
with varying coverage of solar-like active regions and cores of
active regions. However, as discussed in Sect.~\ref{sec3.1}, some
observations suggest a significant flare-like activity that cannot
be described in terms of only active regions and cores. The most
evident case is the June 2002 observation characterized by a
luminosity $3\times$ higher than during all the three maxima in the
coronal cycle (see Fig.~\ref{fig1}). Hints of flare-like activity
are also present in other observations (e.g. April 2011 and April
2015).

Several authors suggested that a superposition of many flare-like
events, ranging in energy from very frequent nano- and micro-flares
to less frequent larger flares, might contribute significantly to
stellar coronal emission (e.g. \citealt{1986Natur.321..679B,
1997ApJ...474..511F, 2000ApJ...545.1074D} and references therein).
In the light of this, here we adopt a similar methodology of Paper
II, but now considering also the contribution of solar-like flaring
regions. The ``quiet Sun'' is omitted because, at its minimum
of coronal activity, the Sun has an X-ray luminosity which is $\sim
2$ orders of magnitude lower than the Sun at maximum (see
\citealt{porrh00} and Fig.~\ref{fig3}) and, in the presence of
active regions, its contribution to X-ray emission is negligible
(see also \citealt{opr01, opr04}). Thus we model the evolution of
the coronal X-ray luminosity and temperature of \hd\ along its cycle
of activity in terms of varying coverage of the stellar surface
with solar-like active regions, cores, and flares. As in Paper II,
our simulations assume the minimum radius, $R=2\,R_{\odot}$, for
the dominant (the primary) component of \hd\, in order to derive
upper limits\footnote{The smaller the star, the higher the filling
factors necessary to reproduce the observed level of emission.} for
the filling factors derived for the various types of coronal
structures.
\begin{figure*}
  \begin{center}
    \leavevmode
        \epsfig{file=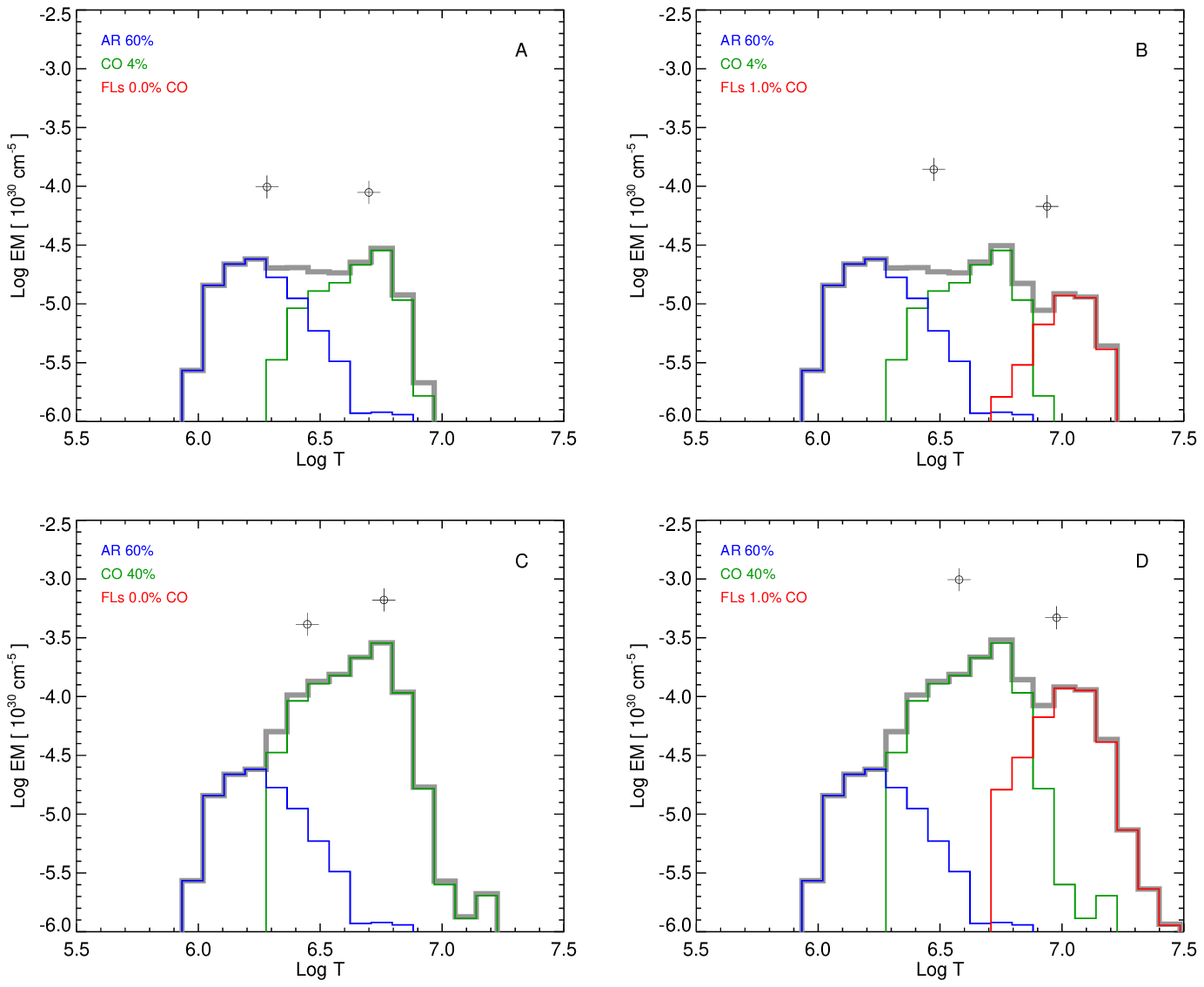, width=17cm}
	\caption{Simulated distributions of emission measure vs.
	temperature, EM($T$), derived for the minimum (upper panels)
	and maximum (lower panels) coverage of cores and for the
	minimum (on the left) and maximum (on the right) coverage
	of flares.  The contributions to EM($T$) of active regions
	(AR, Blue), cores (CO, green), and flares (FLS, red) are
	shown. We assume a constant surface filling factor of active
	regions ($F_{AR} \approx 60$\%), a filling factor of cores
	ranging between 4 (upper panels) and 40\% (lower panels),
	a filling factor of flares ranging between 0 (right panels)
	and 1\% (left panels) of the coverage of cores. The crosses
	indicate the values of temperature and emission measure of
	the two isothermal components fitting the corresponding
	spectra.}
  \label{fig2} \end{center}
\end{figure*}

As a first step in our approach, we consider the average
distribution of emission measure vs. temperature, EM($T$), per unit
surface area for each class of solar coronal structures considered
here (namely active regions, cores, and flares). To this end, we
take advantage of the results of the analysis of observations
collected with the soft X-ray telescope (SXT) on board the Yohkoh
satellite (\citealt{1991SoPh..136...37T}) which are discussed in
previous studies (\citealt{opr01, opr04, 2001ApJ...557..906R}). For
the reader convenience, we summarize here the method followed by
those authors to derive the EM($T$) distributions of specific coronal
regions. In the case of the solar corona outside evident flares,
\cite{opr01} have shown that three different classes of coronal
regions contributing to the X-ray emission of the whole solar corona
can be identified from the analysis of Yohkoh/SXT images: quiet
regions, active regions, and active region cores. The selection of
these regions is based on the SXT pixel intensity (or, in other
words, on the surface brightness), a criterion similar to that
adopted by \cite{1996PhDT........62H} and \cite{2000ApJ...545.1074D}.
Once the regions are selected, the EM($T$) distribution is derived
from the analysis of the selected pixels following the method
outlined by \cite{opr00}. It is worth noting that studies conducted
during a complete solar coronal cycle have shown that the latter
can be explained by a varying surface coverage of active regions
and cores (\citealt{opr01}) a mechanism which we wish to test for
\hd. More specifically, these studies have shown  that the average
temperature of the different classes of regions increases from quiet
regions ($\sim 1$~MK) to active regions ($\sim 2$~MK) to cores
($\sim 5$~MK) (\citealt{opr01}). The contribution of quiet regions
to the X-ray emission does not change appreciably, whereas the
contribution from active regions and cores decreases by several
orders of magnitude with the solar cycle. Indeed although active
regions and cores in the Sun cover only a minor fraction of the
solar surface ($\sim 30$\% at the maximum and $\sim 0.1$\% at the
minimum) they dominate the EM($T$) distribution of the whole solar
corona for temperatures above $2$~MK (outside flares)
and determine the changes in X-ray luminosity and average temperature
of the whole solar corona (\citealt{opr01}). This behaviour justifies
our choice to neglect the contribution of the quiet Sun from our
analysis. For the purposes of the present paper, we derive the
average EM($T$) distributions of active regions and cores, by
considering the analysis of \cite{opr04} who studied the long term
evolution of a solar active region and its core observed with
Yohkoh/SXT from its emergence (July 5, 1996) to the decay phase
(end of October 1996). The average distributions are derived by
averaging the distributions obtained by these authors in different
phases of evolution of these regions.

For the EM($T$) distribution of flares, we consider a sample of
flares observed with Yohkoh/SXT and analyzed by \cite{2001ApJ...557..906R}.
The sample includes eigth flares from weak (GOES class C5.8) to
very intense ones (X9). For each flare, the EM($T$) distribution
of the flaring region was derived by \cite{2001ApJ...557..906R}
in the rise, peak, and decay phase. Here we consider those
distributions derived for all the flares and in different phases
of their evolution and determine the average contribution of flaring
regions to the EM($T$) distribution of the whole corona, by assuming
that the differential energy frequency distribution of solar flares
is described as $N(E)\propto E^{−\alpha} dE$ where $E$ is the flare
energy and the index $\alpha = 1.53$ (e.g.  \citealt{2002ApJ...572.1048A,
2013NatPh...9..465W} and references therein). Then, from the EM($T$)
distributions per unit surface area for each type of solar coronal
structures, we explore different surface coverage of active regions,
cores, and flares and derive the resulting EM($T$) distribution of
the whole corona and the corresponding \xmm\ spectrum and X-ray
luminosity.  Finally we compare the simulated data with those of
\hd.

In Paper II, we found that the coronal emission is largely dominated
by cores out of the cycle minimum so that the coverage of active
regions can be constrained only during the minimum. Following the
results of Paper II, the minimum in 2005 can be described by assuming
a coverage of active regions with a maximum surface filling factor
of about $F_{AR} \approx 60$\% (corresponding to 27\% if considering
the upper limit to the star radius, $R_1 = 3\,R_{\odot}$) and a
coverage of cores with $F_{CO} \approx 4$\%. We assume, therefore,
the same coverage of active regions during the whole evolution.
This assumption reflects our unability to constrain the active
regions coverage. However it is unlikely that the active regions
coverage does not vary during the cycle as shown, for instance, in
our Sun. In Sect.~\ref{scaledsun} we discuss this point in more
detail. With the above approximation, the cyclic modulation of
coronal activity is determined by varying only the coverage of cores
and flares.  We allow the filling factor of cores, $F_{CO}$, to
vary between 4 and 40\%, the latter corresponding to a full coverage
of the stellar surface considering the filling factor of active
regions at 60\% and a stellar radius of $2\,R_{\odot}$. We assume
the coverage of flares to depend on the coverage of cores and allow
the filling factor of flares, $F_{FLs}$, to vary between 0 and 1\%
of the coverage of cores (in other words, we allow a small percentage
of the cores to be flaring).  Figure~\ref{fig2} shows the modeled
EM($T$) distributions for the minimum and maximum allowed coverage
of cores and flares. As expected, the contribution from active
regions to the EM($T$) distribution is significant only when the
coverage of cores is minimum (upper panels in Fig.~\ref{fig2}).  In
that case the contribution of active regions is comparable to those
of cores.  Conversely, when the coverage of cores is the maximum
(40\%), the EM($T$) distribution is largely dominated by cores. As
a result, the surface filling factor of active regions cannot be
univocally determined. The flares contribute to the high temperature
portion of the EM($T$) distribution, leading to an higher average
coronal temperature and to an harder spectrum.

\begin{figure}
  \begin{center} 
    \leavevmode 
        \epsfig{file=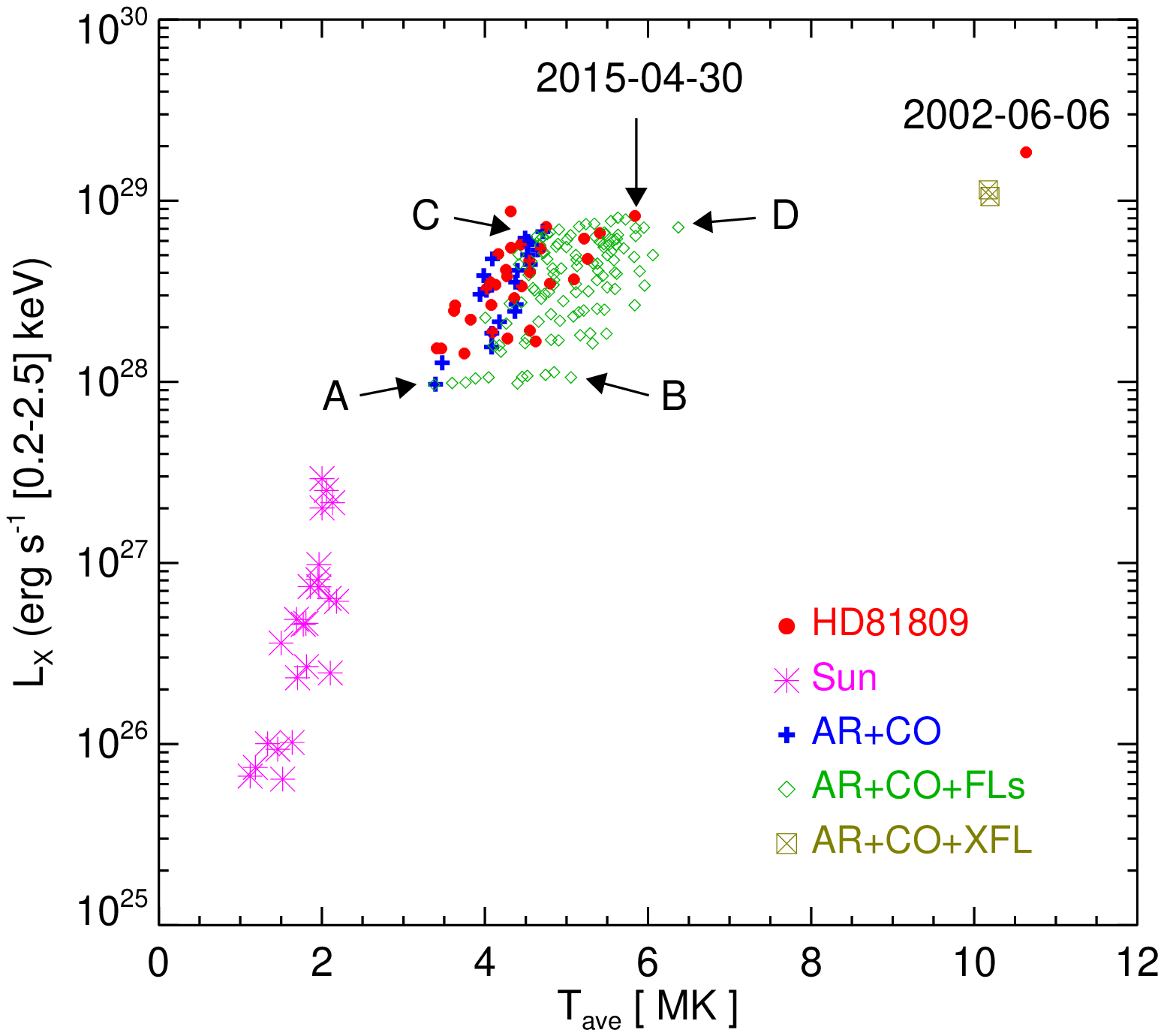, width=9.cm}
	\caption{Evolution of coronal X-ray temperature and luminosity
	along the cycle for the Sun (magenta stars; data adapted
	from \citealt{opr01}) and \hd\ (red filled circles). The
	figure also reports the simulated data obtained assuming a
	mixture of solar-like coronal regions: active regions and
	cores (namely non-flaring regions; AR+CO, blue crosses),
	and active regions, cores, and flares (AR+CO+FLs, green
	diamonds). The crossed square (AR+CO+XFL) marks the synthetic
	values considering the contribution of a very intense flare
	(GOES class X9) to the non-flaring corona with maximum
	coverage of cores ($\sim 40$\%). Labels A-D mark the simulated
	data points corresponding to the EM($T$) distributions in
	Fig.~\ref{fig2}. The data from actual spectra of June 2002
	and April 2015 are indicated.}
  \label{fig3}
  \end{center}
\end{figure}

From the EM($T$) distributions we synthesize the corresponding
\xmm\ spectra and analyze them in the same way as the actual data
of \hd, using XSPEC. In particular we estimate the parameters
characterizing the simulated spectra (namely temperature and emission
measure) by fitting the spectra with two isothermal components (APEC
models). Then we derive the coronal luminosity, $L_{X}$, and the
emission measure-weighted temperature, $T_{ave}$, defined in
Eq.~\ref{tave}. Figure~\ref{fig3} reports the range of variation
in $L_{X}$ and $T_{ave}$ obtained for the simulated data (blue and
green symbols) using the above approach, together with those observed
for the Sun (magenta symbols; data adapted from \citealt{opr01})
and \hd\ (red symbols) along their cycle. The figure clearly shows
that the simulated data cover the entire observed range of variations
of \hd\ in terms of both $L_{X}$ and $T_{ave}$, so that the entire
cycle of \hd\ can be reproduced by simply assuming the same coverage
of active regions ($F_{AR} = 60$\%) and a different coverage of
cores and flares.  Both coronal temperature and luminosity at the
cycle minimum can be explained if the active regions cover $\approx
60$\% of the stellar surface, the cores about 4\%, and there are
no flaring regions, in agreement with Paper II. An EM($T$) distribution
close to that at cycle minimum is shown in panel A of Fig.~\ref{fig2}
(see also the corresponding data point labeled A in Fig.~\ref{fig3}).
For comparison, the average temperature and the X-ray luminosity
of the solar corona at minimum are the result of active regions
covering no more than 0.1\% of the solar surface, with no cores and
flares present (\citealt{opr01}). In fact, the coronal temperature
of the Sun at minimum is a factor of 3 lower and the X-ray luminosity
two orders of magnitude lower than those of \hd\ at minimum (see
Fig.~\ref{fig3}). This is due in part also to the smaller radius
of the Sun with respect to that of the primary component of \hd.
In fact, in Paper II we found that the X-ray surface flux (in the
$[0.2-2.5]$~keV band) of the Sun at minimum is only a factor of
$\approx 4$ lower than that of \hd\ at minimum (see Fig.~3 in Paper
II).

The coronal luminosity of \hd\ close to the cycle maximum can be
reproduced if the active regions cover $\approx 60$\% of the stellar surface
and the cores about 40\% (panel C in Fig.~\ref{fig2} and data point
labeled C in Fig.~\ref{fig3}). However, out of the minimum, some
of the data points show values of temperature that cannot be simply
explained as a mixture of active regions and cores only (blue symbols in
Fig.~\ref{fig3}). In these cases a contribution of flares is required
(green symbols in Fig.~\ref{fig3}). For instance the April 2015
data (see Fig.~\ref{fig3}), characterized by a coronal temperature
$T_{ave} \sim 5.8$~MK can be represented if, for instance, the active regions
cover $\approx 60$\% of the stellar surface, the cores about 40\%, and
1\% of cores is flaring (data point labeled D in Fig.~\ref{fig3}). The
corresponding EM($T$) distribution is shown in panel D of
Fig.~\ref{fig2}. In the case of the Sun, at solar maximum the coronal
temperature and luminosity are determined by a surface coverage of
active regions of about 30\%, a coverage of cores of about 1\%, and
no flares present (\citealt{opr01}). As a result, the coronal
temperature and luminosity of the Sun at maximum are still significantly
lower that those of \hd\ at minimum (see Fig.~\ref{fig3}). Again
the different luminosity is partially due to the different radius
of the two stars; in fact the X-ray surface flux of the Sun at
maximum is only slightly lower than that of \hd\ at maximum (see
Fig.~3 in Paper II). This cannot explain however the different
average temperature. We note that the less massive companion in
\hd, with a mass $M_2 = 1.0 \pm 0.25$\,M$_{\odot}$ very close to
that of the Sun, has most likely levels of activity similar to those
of the Sun. Its luminosity, therefore, is expected to be much lower
than that of the more massive larger companion, even at maximum of
activity. This justifies our assumption that the X-ray activity of
the system is dominated by the primary component.

The above results suggest that, besides the June 2002 data set (with a
coronal temperature $T_{ave} \sim 11$~MK in Fig.~\ref{fig3}) which
is clearly an outlier separated from the cloud of values of \hd,
other data points in Fig.~\ref{fig3} might have an unresolved flaring
component contributing to the coronal emission. In Paper II, we
interpreted the June 2002 data point as a flare-like phenomenon.
This is supported by the evidence that its spectral characteristics
can be reproduced if the contribution of a very intense solar flare
(a GOES class X9 flare observed with Yohkoh/SXT on November 2, 1992)
is added to the synthetic X-ray spectrum simulating the cycle maximum
of \hd\ (see crossed square in Fig.~\ref{fig3}). We wonder, therefore,
if the systematic higher coronal temperature of some of the data
points with respect to simulations considering only a mixture of
active regions and cores can be considered as evidence of flares not visible as
short-term variability in the \xmm\ observations. As shown by
\cite{opr04} in their study of the Sun as a star, the characteristics
of X-ray variability due to the coronal cycle and to flaring activity
are different. A prominent difference is that the variability induced
by flares leads to a significant change of coronal temperature,
whereas the variability due to the cycle is mainly characterized
by large variations in X-ray luminosity (see Fig.~9 in \citealt{opr04}).
This is evident also in Fig.~\ref{fig3} where the variability induced
by an increased flaring activity (for instance, evolving from panel
A to panel B in Fig.~\ref{fig2}) leads to an evolution from data
point A to data point B; the variability induced by an increasing
coverage of cores (namely evolving from panel A to panel C in
Fig.~\ref{fig2}) leads to an evolution from data point A to data
point C in Fig.~\ref{fig3}. In other words, flaring regions
significantly increase the average coronal temperature, a feature
that seems to characterize some of the data points of \hd\ (e.g.
the April 2015 data, see Fig.~\ref{fig3}).

\begin{figure}
  \begin{center}
    \leavevmode
        \epsfig{file=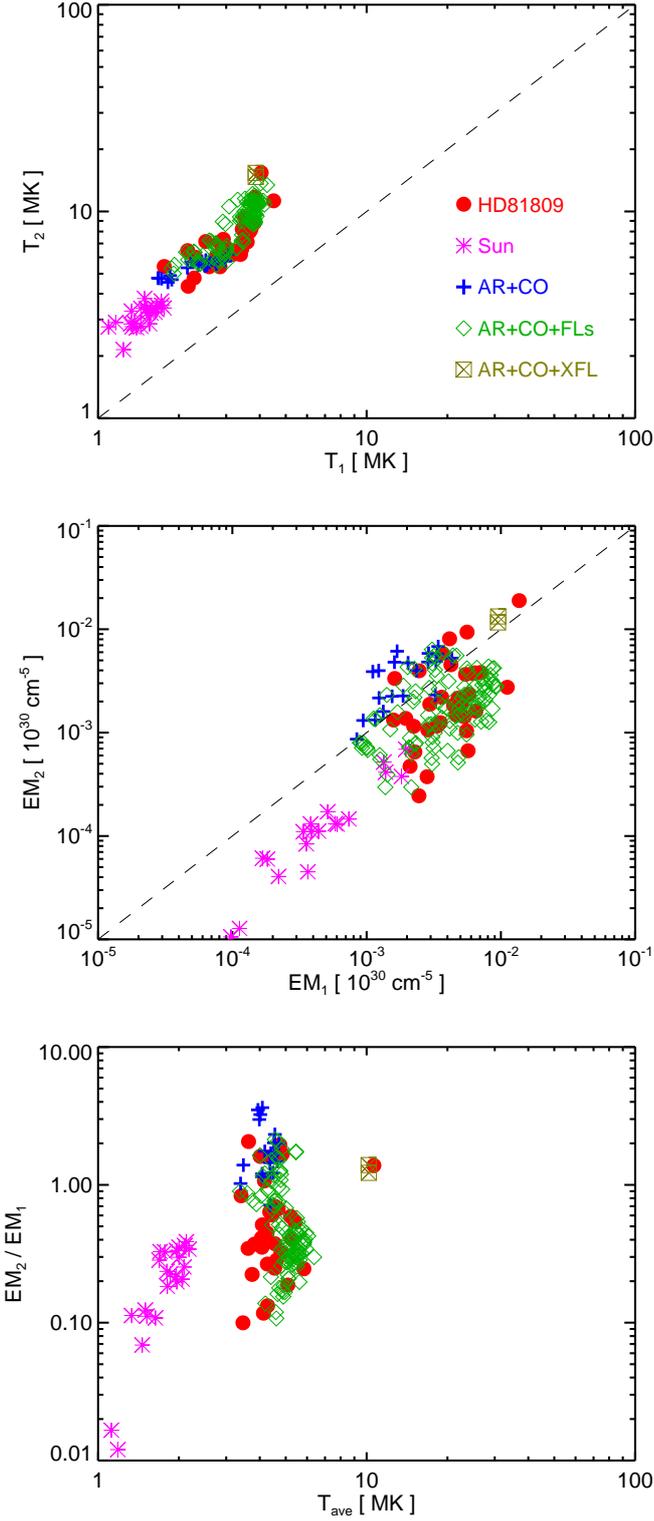, width=9.cm}
	\caption{Parameters derived from the fitting with two
	isothermal components of the actual spectra of \hd\ (red
	circles) and of the Sun observed as a star (magenta stars),
	along their cycles. The figure also shows the fitting
	parameters derived for the simulated data assuming a mixture
	of solar-like coronal regions: active regions and cores
	(AR+CO, blue crosses), and active regions, cores, and flares
	(AR+CO+FLs, green diamonds). The crossed square (AR+CO+XFL)
	marks a mixture of active regions and cores plus the
	contribution of a very intense flare (GOES class X9).}
  \label{fig4}
  \end{center}
\end{figure}

In Fig.~\ref{fig4} we compare the parameters derived from the
spectral fitting of the actual data of \hd\ with those derived from
the fitting of simulated spectra (colors and symbols as in
Fig.~\ref{fig3}). Again we find that the range of variations of the
parameters characterizing the spectra of \hd\ along its cycle are
well represented in terms of a mixture of active regions, cores, and flares (green
diamonds in Fig.~\ref{fig4}). Here the comparison of fitting parameters
derived from observed and simulated spectra shows clearly that
a mixture of only active regions and cores (blue crosses in Fig.~\ref{fig4}) cannot
describe most of the data points of \hd. In fact, in many
cases the observed spectra require the second isothermal component
with higher values of $T_2$ and/or lower values of EM$_2$ (and ratio
EM$_2$/EM$_1$ lower than 1) than the spectra simulated with a mixture
of active regions and cores. We conclude that adding a contribution from flares is
necessary to reproduce the spectral characteristics of all data
points of \hd. Again this result supports the scenario of an
unresolved significant flaring activity in \hd. This is not evident
if the observed spectra are fit with one isothermal component, as
we did in Paper II.

\subsection{A scaled up Sun?}
\label{scaledsun}

To probe the distribution of coronal temperature in \hd\ during its
cycle and to attempt a rough reconstruction of its EM($T$) distribution,
a three-temperatures model was adopted, in which the temperatures
were fixed at $T_1 = 0.17$~keV ($\sim 2$~MK), $T_2 = 0.43$ keV
($\sim 5$~MK), and $T_3 = 1.09$~keV ($\sim 13$~MK), namely the
values roughly corresponding to the peaks of emission measure of
the EM($T$) distributions of active regions, cores, and flares, respectively (see
Fig.~\ref{fig2}). For this model, therefore, only the normalizations
(emission measures) were allowed to vary. Table~\ref{tab2} reports
the best-fit parameters from the three-temperatures model, for all
the data set of \hd.

\begin{table}
  \begin{center}
    \caption{Best-fit spectral parameters from the three-temperatures
    ($\sim$ 2, 5, 13~MK) model, for the 33 \xmm\ observations of \hd.}
    \begin{tabular}{ccccc}
\hline\hline
Date & EM$_1$ & EM$_2$ & EM$_3$ \\
     & [$10^{27}$~cm$^{-5}$] & [$10^{27}$~cm$^{-5}$] & [$10^{27}$~cm$^{-5}$] \\\hline
\vspace{-0.25cm} \\
2001-11-01  &  $2.01^{+0.81}_{-0.81}$ & $6.71^{+0.56}_{-0.56}$ & $1.43^{+0.56}_{-0.34}$ \\
\vspace{-0.25cm} \\
2002-06-06  &  $7.16^{+1.33}_{-1.33}$ & $6.77^{+0.81}_{-0.81}$ & $14.0^{+0.81}_{-0.65}$ \\
\vspace{-0.25cm} \\
2002-11-02  &  $3.67^{+1.26}_{-1.26}$ & $5.40^{+0.76}_{-0.76}$ & $0.00^{+0.76}_{-0.00}$ \\
\vspace{-0.25cm} \\
2003-05-03  &  $1.20^{+0.59}_{-0.59}$ & $3.93^{+0.39}_{-0.39}$ & $0.30^{+0.39}_{-0.23}$ \\
\vspace{-0.25cm} \\
2003-11-22  &  $2.30^{+0.36}_{-0.36}$ & $2.07^{+0.23}_{-0.23}$ & $0.00^{+0.23}_{-0.00}$ \\
\vspace{-0.25cm} \\
2004-04-29  &  $1.42^{+0.14}_{-0.14}$ & $1.29^{+0.09}_{-0.09}$ & $0.00^{+0.09}_{-0.00}$ \\
\vspace{-0.25cm} \\
2004-11-09  &  $0.87^{+0.15}_{-0.15}$ & $1.90^{+0.11}_{-0.11}$ & $0.00^{+0.11}_{-0.00}$ \\
\vspace{-0.25cm} \\
2005-05-03  &  $1.07^{+0.14}_{-0.14}$ & $1.37^{+0.09}_{-0.09}$ & $0.00^{+0.09}_{-0.00}$ \\
\vspace{-0.25cm} \\
2005-10-30  &  $0.69^{+0.11}_{-0.11}$ & $1.89^{+0.08}_{-0.08}$ & $0.07^{+0.08}_{-0.05}$ \\
\vspace{-0.25cm} \\
2006-11-05  &  $1.68^{+0.21}_{-0.21}$ & $4.66^{+0.15}_{-0.15}$ & $0.00^{+0.15}_{-0.00}$ \\
\vspace{-0.25cm} \\
2007-05-04  &  $1.26^{+0.24}_{-0.24}$ & $5.19^{+0.18}_{-0.18}$ & $0.81^{+0.18}_{-0.11}$ \\
\vspace{-0.25cm} \\
2007-11-09  &  $2.20^{+0.25}_{-0.25}$ & $4.09^{+0.17}_{-0.17}$ & $0.00^{+0.17}_{-0.00}$ \\
\vspace{-0.25cm} \\
2008-05-14  &  $2.25^{+0.30}_{-0.30}$ & $6.18^{+0.21}_{-0.21}$ & $1.17^{+0.21}_{-0.13}$ \\
\vspace{-0.25cm} \\
2008-11-06  &  $2.57^{+0.29}_{-0.29}$ & $6.27^{+0.21}_{-0.21}$ & $0.00^{+0.21}_{-0.00}$ \\
\vspace{-0.25cm} \\
2009-04-27  &  $2.64^{+0.28}_{-0.28}$ & $5.61^{+0.20}_{-0.20}$ & $0.00^{+0.20}_{-0.00}$ \\
\vspace{-0.25cm} \\
2009-05-24  &  $1.93^{+0.32}_{-0.32}$ & $4.74^{+0.22}_{-0.22}$ & $0.00^{+0.22}_{-0.00}$ \\
\vspace{-0.25cm} \\
2009-11-04  &  $2.46^{+0.30}_{-0.30}$ & $3.52^{+0.20}_{-0.20}$ & $0.00^{+0.20}_{-0.00}$ \\
\vspace{-0.25cm} \\
2009-12-08  &  $2.26^{+0.31}_{-0.31}$ & $4.97^{+0.22}_{-0.22}$ & $0.27^{+0.22}_{-0.13}$ \\
\vspace{-0.25cm} \\
2010-04-30  &  $2.27^{+0.29}_{-0.29}$ & $3.36^{+0.20}_{-0.20}$ & $0.00^{+0.20}_{-0.00}$ \\
\vspace{-0.25cm} \\
2010-10-31  &  $1.55^{+0.26}_{-0.26}$ & $3.82^{+0.18}_{-0.18}$ & $0.00^{+0.18}_{-0.00}$ \\
\vspace{-0.25cm} \\
2011-04-27  &  $1.69^{+0.31}_{-0.31}$ & $3.54^{+0.21}_{-0.21}$ & $0.62^{+0.21}_{-0.13}$ \\
\vspace{-0.25cm} \\
2011-11-02  &  $1.73^{+0.23}_{-0.23}$ & $2.06^{+0.15}_{-0.15}$ & $0.00^{+0.15}_{-0.00}$ \\
\vspace{-0.25cm} \\
2012-05-07  &  $1.98^{+0.25}_{-0.25}$ & $2.55^{+0.16}_{-0.16}$ & $0.00^{+0.16}_{-0.00}$ \\
\vspace{-0.25cm} \\
2012-06-07  &  $1.85^{+0.26}_{-0.26}$ & $2.62^{+0.17}_{-0.17}$ & $0.00^{+0.17}_{-0.00}$ \\
\vspace{-0.25cm} \\
2012-11-01  &  $1.14^{+0.25}_{-0.25}$ & $1.46^{+0.16}_{-0.16}$ & $0.00^{+0.16}_{-0.00}$ \\
\vspace{-0.25cm} \\
2013-04-27  &  $0.99^{+0.18}_{-0.18}$ & $2.00^{+0.12}_{-0.12}$ & $0.11^{+0.12}_{-0.07}$ \\
\vspace{-0.25cm} \\
2013-10-30  &  $1.59^{+0.22}_{-0.22}$ & $1.51^{+0.13}_{-0.13}$ & $0.16^{+0.13}_{-0.08}$ \\
\vspace{-0.25cm} \\
2014-04-28  &  $1.14^{+0.38}_{-0.38}$ & $3.44^{+0.26}_{-0.26}$ & $0.00^{+0.26}_{-0.00}$ \\
\vspace{-0.25cm} \\
2014-10-29  &  $1.42^{+0.26}_{-0.26}$ & $3.79^{+0.18}_{-0.18}$ & $0.00^{+0.18}_{-0.00}$ \\
\vspace{-0.25cm} \\
2015-04-30  &  $1.08^{+0.43}_{-0.43}$ & $9.16^{+0.33}_{-0.33}$ & $1.87^{+0.33}_{-0.21}$ \\
\vspace{-0.25cm} \\
2015-10-29  &  $2.09^{+0.34}_{-0.34}$ & $6.39^{+0.24}_{-0.24}$ & $0.00^{+0.24}_{-0.00}$ \\
\vspace{-0.25cm} \\
2016-04-25  &  $2.24^{+0.38}_{-0.38}$ & $8.52^{+0.27}_{-0.27}$ & $0.33^{+0.27}_{-0.15}$ \\
\vspace{-0.25cm} \\
2016-10-29  &  $2.84^{+0.57}_{-0.57}$ & $9.84^{+0.41}_{-0.41}$ & $0.84^{+0.41}_{-0.24}$ \\
\vspace{-0.25cm} \\
\hline
    \end{tabular}
    \label{tab2}
  \end{center}
\end{table}

\begin{figure}
  \begin{center}
    \leavevmode
        \epsfig{file=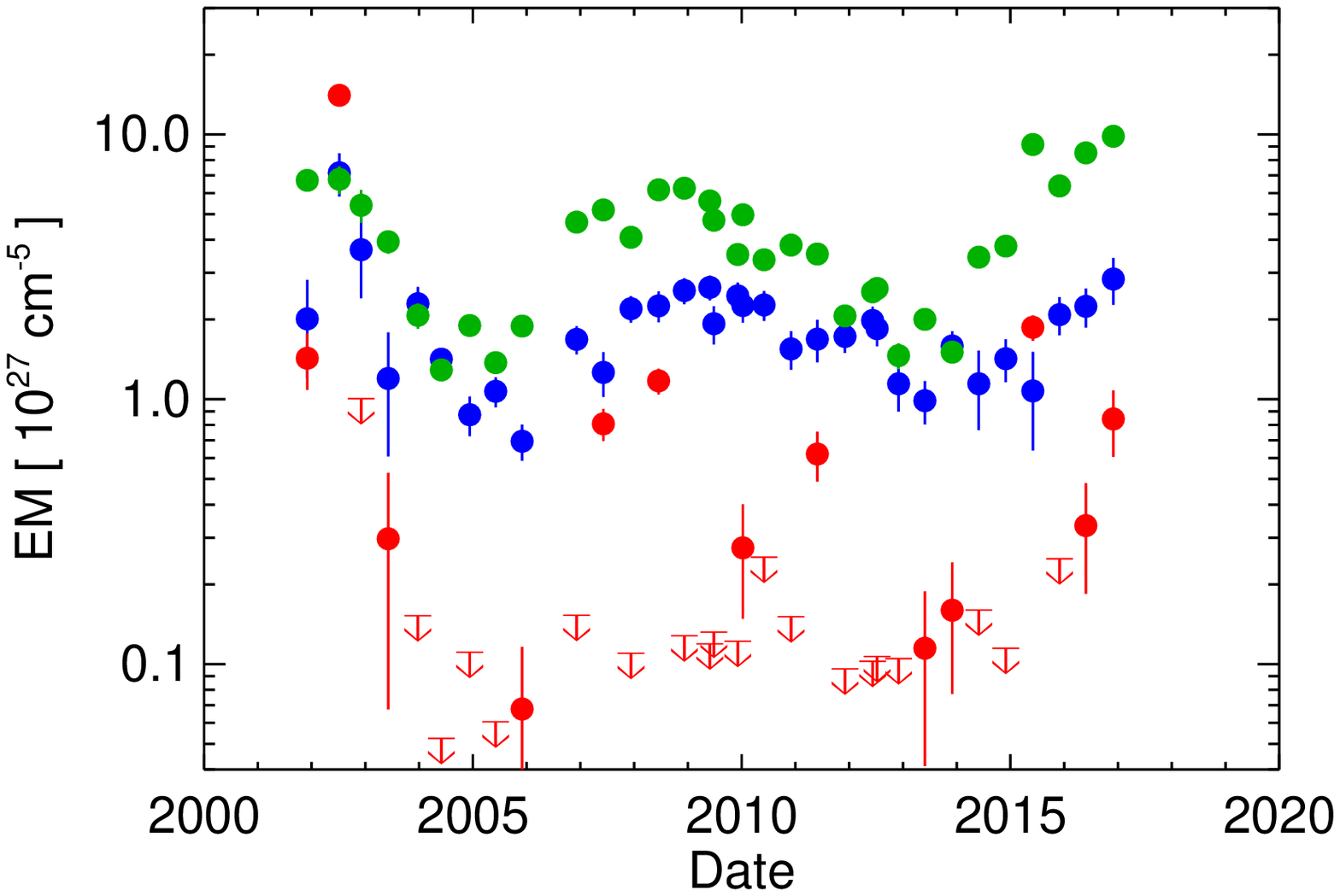, width=9.cm}
        \epsfig{file=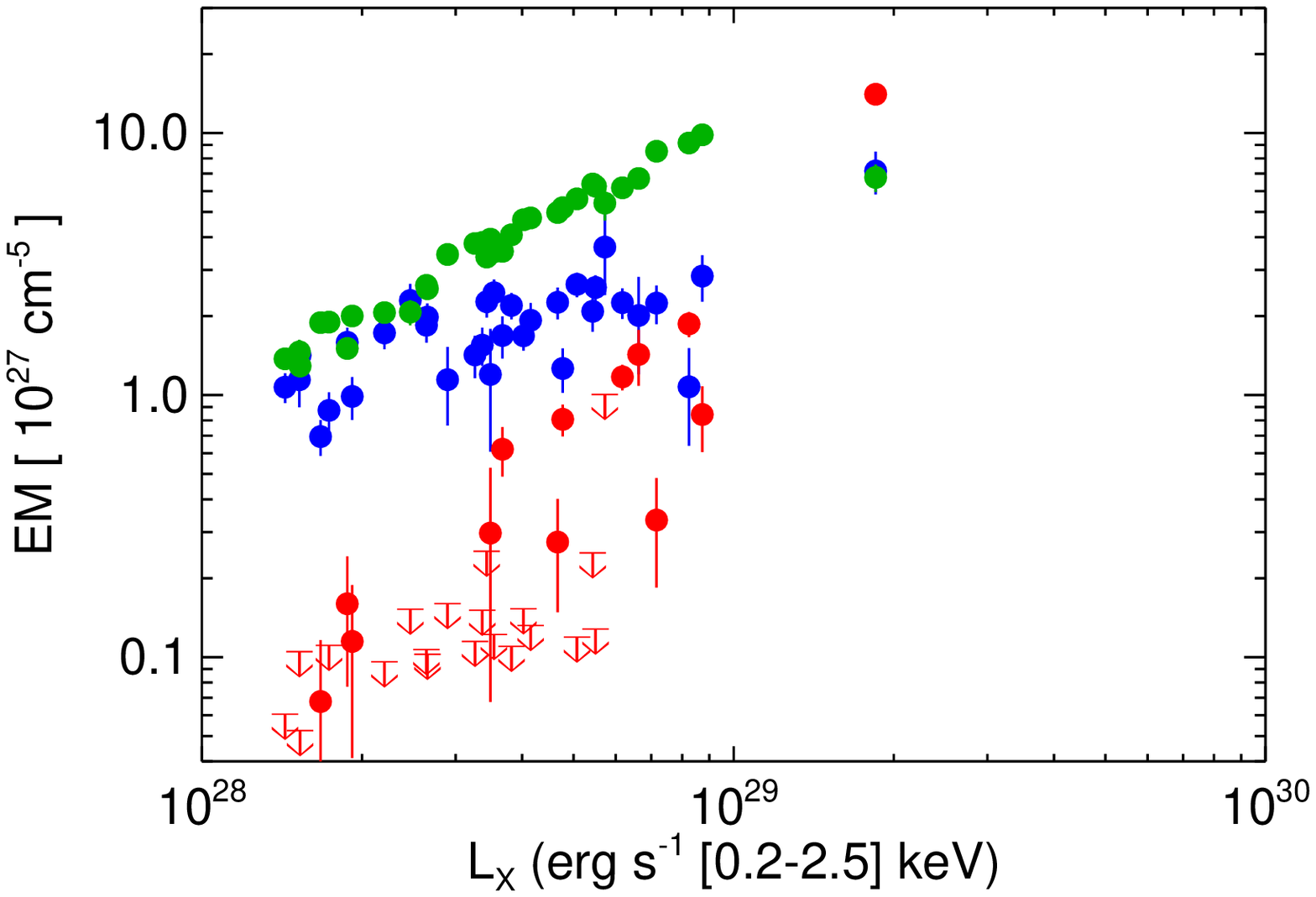, width=9.cm}
	\caption{Best-fit emission measures from the three-temperatures
	model along the coronal cycle of \hd\ (upper panel) and as
	a function of coronal luminosity (lower panel). The figure
	shows the parameters for the coldest (blue circles,
	corresponding to active regions), intermediate (green, corresponding
	to cores), and hottest (red, corresponding to flares) components.
	Upper limits are marked with arrows.}
  \label{fig5} \end{center}
\end{figure}

The variability of the best-fit emission measures along the coronal
cycle of \hd\ is shown in Fig.~\ref{fig5}. The cycle modulation is
particularly evident in the emission measure of the intermediate
(green circles) and coldest (blue) isothermal components (upper
panel in the figure). The hottest (red circles) component shows
many upper limits to the emission measure and a weaker relation
with the coronal cycle; nevertheless, for this component, the highest
values of EM occur, in general, in proximity of cycle maxima. If
we interpret the isothermal components in terms of solar-like coronal
regions, the cold component corresponds to the active regions, the intermediate
to the cores, and the hot to the flares.

This new approach in the analysis of actual spectra confirms that
the cyclic modulation is mainly due to the varying coverage of cores.
The lower panel in Fig.~\ref{fig5} shows a clear correlation between
the emission measure of cores and the coronal luminosity of \hd\ with
the EM increasing by about one order of magnitude between the minimum
and maximum of the cycle. We note that this is compatible with the
increase of $F_{CO}$ from 4 to 40\% in our simulations.  Thus, we
speculate that this regular and gradual increase reflects the
increase of surface filling factor of cores. The figure shows a
correlation also for the active regions but with the EM varying only by a factor
$\approx 3$ excluding the June 2002 data point. The EM of active regions is
comparable to that of cores close to the minimum; then the EM of cores
increases with the level of activity faster than that of active regions and it
is a factor of $\sim 3$ larger than that of active regions close to the maximum.
This result shows that the coverage of active regions cannot be considered
permanent as assumed in Sect.~\ref{sec3.2}; a varying coverage of
active regions contributes to the cyclic modulation of coronal activity. This
was expected also considering that, in the Sun, the coverage of active regions varies
during the coronal cycle. In fact our choice to consider a permanent
coverage of active regions was dictated by our unability to constrain the active regions
coverage. In addition it is worth to note that the adopted filling
factor, $F_{AR} = 60$\%, is an upper limit when a star radius $R_1 =
2\,R_{\odot}$ is considered. For $R_1 = 3\,R_{\odot}$, the upper limit
is even lower, $F_{AR} \approx 27$\%, thus allowing a variation of active regions coverage
during the cycle of a factor of 3, as inferred from Fig.~\ref{fig5}.
Despite the increasing active regions coverage toward the maximum, Fig.~\ref{fig5}
confirms that the active regions contribution is not determinant for the observed
variability which is, instead, mostly due to the varying coverage
of cores. The contribution of flares is, in general, the lowest in terms
of EM, although it increases rapidly with the level of coronal
activity, becoming comparable to the contribution of active regions in proximity
of cycle maxima. This means that their contribution to the X-ray
luminosity is not determinant as also shown by Fig.~\ref{fig3}.
Nevertheless the hot isothermal component (or, in other words, the
flares) is important to fit the high energy part of the spectrum, as
demonstrated by the fitting with two isothermal components of actual
spectra (see Fig.~\ref{fig4}) and by the average coronal X-ray
temperature (see Table~\ref{tab1} and Fig.~\ref{fig3}).

Figure~\ref{fig5} also shows that the June 2002 observation differs
significantly from the others. The coronal emission is largely
dominated by the hot component (flares) which has the largest value
of EM. The other two components (active regions and cores) have similar EM, a
factor of $\approx 2$ lower than that of the hot component (see
Table~\ref{tab2}). This result supports the idea that a bright flare
was dominating the coronal emission of \hd\ in the June 2002 data.
Interestingly, in this case, the EM of the hot component rather
than that of the intermediate one seems to be an extension of the
distribution of data points of the cores (see the lower panel in
Fig.~\ref{fig5}). This reinforces the peculiarity of this data point
with respect to the others.

The scenario suggested by the fitting with a three-temperatures
model is, in some way, analogous to that discussed by
\cite{2000ApJ...545.1074D} to explain the high levels of coronal
emission measure distributions derived for the intermediate-activity
stars $\epsilon$ Eri (K2 V) and $\xi$ Boo A (G8 V) from Extreme
Ultraviolet Explorer spectroscopic observations. These authors found
that the coronae of solar-like stars with activity levels up to
that of $\xi$ Boo A can be explained in terms of coronal structures
(active regions and cores of active regions) of the same type of
that observed in the Sun. The activity level follows the surface
area coverage of these regions: in higher activity stars the coverage
approaches unity. We observe a similar behavior in \hd\, where
its activity level during the coronal cycle is modulated by
the varying coverage of active regions and cores which is the maximum (around
unity) in correspondence of cycle maxima. \cite{2000ApJ...545.1074D}
suggest that, as the surface area coverage of active regions and cores is close
to unity, the interactions and merging of active coronal regions
should increase, leading to a corresponding increase of flaring activity
in terms of frequency and intensity of flares. This is consistent
with the evidence of a significant (unresolved) flaring activity
in the data of \hd\, especially in proximity of its cycle maxima.
The best evidence of a bright flare is in the June 2002 data point, which
shows the highest X-ray luminosity of \hd, close to the first cycle
maximum, recorded in the long-term monitoring program of \xmm. We
conjecture that the increasing surface coverage of active regions and cores in
\hd\, when it approaches a cycle maximum, makes the interaction
among these regions more effective. As a result, higher frequency
and larger flares could be produced than would be within individual
active regions. We suggest that an unresolved flaring activity in \hd\
is responsible for heating plasma to coronal temperatures of $T
\approx 10$~MK, explaining the evidence of a hot plasma component
in some spectra of \hd\ especially close to the cycle maxima.

\section{Summary and Conclusions}
\label{sec4}

We have presented the analysis of the complete data set resulting
from the long-term \xmm\ monitoring program of the solar-like star
\hd, covering 15 years of evolution. The star is characterized by
a well-defined cycle in its chromospheric activity. We have shown
here that \hd\ is also characterized by a cyclic variability of
coronal X-ray luminosity which is coherent with the chromospheric
one. The data covers approximately two cycles of coronal activity;
the modulation of X-ray luminosity has an amplitude of a factor of
$\sim 5$ (excluding an evident flare in the June 2002 observation)
and a period of $7.3\pm 1.5$~yr, consistent with that of the
chromospheric cycle. The two coronal cycles are very similar,
pointing at a regular modulation of the coronal activity.

\cite{2003ApJ...598.1387P} studied the relationship between the
total magnetic flux and the X-ray luminosity inferred from X-ray
and magnetic field observations of the Sun and of a sample of active
stars (including dwarfs and pre-main-sequence stars). They found
that the magnetic flux and X-ray luminosity exhibit a linear
relationship over 12 orders of magnitude, suggesting an universal
relationship between magnetic flux and the power dissipated through
coronal heating. This relationship can be particularly useful when
magnetic measurements are unavailable and one can use X-rays as a
proxy for the magnetic flux. In the case of \hd, the observed
variation in X-ray luminosity suggests that the magnetic flux ranges
between $3\times 10^{24}$~G~cm$^{2}$ at minimum and $2\times
10^{25}$~G~cm$^{2}$ at maximum. These values are significantly
larger than those measured in the Sun and ranging between $\approx
4\times 10^{22}$~G~cm$^{2}$ at minimum and $\approx 3\times
10^{23}$~G~cm$^{2}$ at maximum.

The spectra of \hd\ have been interpreted in terms of a mixture of
solar-like coronal structures (active regions, cores of active
regions, and flaring regions). We have found that the observed range
of variations of \hd\ along the cycle, in terms of coronal temperature
and luminosity, can be explained with a different surface coverage
of the various types of solar coronal structures, suggesting that
the corona of \hd\ can be considered as a scaled-up version of the
solar corona. The activity level of \hd\ is mainly regulated by a
varying coverage of the cores. At the minimum of activity, our
analysis suggests that the EM($T$) distribution of \hd\ is determined
by comparable contributions from active regions and cores.  Assuming
the lower limit to the radius of \hd\ (namely $R_1 = 2\,R_{\odot}$),
we have found that, to describe the minimum, the surface coverage
of active regions should be $\approx 60$\% and that of cores $\approx
4$\%, with no flares present.  This coverage is much larger than
that observed in the Sun at its maximum of activity and, in addition,
for a star with a much larger radius than the Sun. This justifies
the higher level of activity of \hd\ and the significantly higher
values of coronal temperature and luminosity at minimum with respect
to those of the Sun at maximum.

When \hd\ increases its level of activity, the coverage of cores
gradually increases from $\approx 4$\% at minimum up to $\approx
40$\% in proximity of the maximum. The EM($T$) distribution becomes
dominated by cores and, in some cases, by flares, making the
contribution from active regions negligible and their surface
coverage not univocally determined.  In the light of this, we assumed
a permanent coverage of active regions during the whole cycle with
an upper limit to the filling factor derived at minimum ($F_{AR}
\approx 60$\%). The spectral analysis based on a three-temperatures
model has confirmed the interpretation that the variation of coronal
activity of \hd\ is mainly regulated by a varying coverage of cores
spanning one order of magnitude. In addition, this analysis has
also allowed us to identify a moderate change (a factor of 3) in
the active regions coverage which follows that of cores during the
coronal cycle. In many cases, mostly close to the maximum, the
actual data are characterized by a coronal temperature higher than
that of simulations assuming a coverage of only active regions and
cores. In these cases a significant contribution of flares is
necessary to fit the data. We have interpreted this result as
evidence of an unresolved significant flaring activity present in
\hd\ which increases for increasing level of activity. Following
\cite{2000ApJ...545.1074D}, we speculate that, as the surface filling
factors of active regions and cores increase and approach unity,
the intensity and frequency of flaring events increase because of
the increasing complexity of the coronal structures which interact
each other.  This is also consistent with the evidence that the
intensity and frequency of flares in the Sun increase with its
coronal cycle.

A comparison of the X-ray cycle of \hd\ with those of other late-type
stars reported in the literature shows that \hd\ has a cycle with
similar characteristics. The cycle of 61 Cygni A (a K5V star with
a mass $M = 0.7\,M_{\odot}$) has been monitored with {\it ROSAT}
and \xmm\ in the period 1993-2012 (covering $\approx 19$~yr of
evolution; \citealt{2003A&A...406L..39H, 2006A&A...460..261H,
2012A&A...543A..84R}). Its cycle shows many similarities with that
of \hd. It is characterized by regular variations of X-ray luminosity
with an amplitude of a factor of $\approx 3$ (slightly lower than
that of \hd) and a period of $\approx 7$~yr, consistent with \caii\
measurements. Another example is $\alpha$ Cen that has been monitored
in X-rays by several instruments ({\it ROSAT}, \xmm, {\it Chandra})
in the period 1995-2013 (covering $\approx 18$~yr of evolution;
\citealt{2005A&A...442..315R, 2009ApJ...696.1931A, 2012A&A...543A..84R,
2014AJ....147...59A, 2017MNRAS.464.3281W}). In this case, the
observations suggest that the B component (K1V with $M = 0.91\,M_{\odot}$)
is characterized by an X-ray cycle with a minimum-to-peak contrast
of 4.5 and a period of $8.1 \pm 0.2$~yr (consistent with the
chromospheric cycle), whereas the A component (G2V with $M =
1.1\,M_{\odot}$) was in a Maunder-Minimum-like low state since 2005
and now is approaching a maximum (\citealt{2016arXiv161206570R})
with a possible period of about $15-20$~yr. The B component,
therefore, shows amplitude and period of X-ray variations remarkably
similar to those found for \hd. Conversely, an X-ray cycle with
rather different characteristics is that of the young ($\approx
600$~Myr) late-type star $\iota$ Hor observed with \xmm\ over 21
months in 2011-2013 (\citealt{2013A&A...553L...6S}). In fact, $\iota$
Hor, at variance with the other cases, shows a somewhat irregular
1.6-yr X-ray cycle very similar to the contemporaneous chromospheric
cycle. Indication of a second longer cycle superposed to the shorter
one comes from the apparent amplitude modulation of the 1.6-yr
cycle. In the period monitored by \xmm, the X-ray flux shows variation
of a factor of $\approx 2$ (lower than the amplitude found for \hd).
Interestingly,
a brief episode of chaotic variability is also present when the
longer cycle seems to approach its maximum. After this episode,
the cyclic behavior resumes. The complex variability revealed for
$\iota$ Hor is most likely due to the youth of the star which makes
a major difference with the other late-type stars for which an X-ray
cycle has been detected.

\begin{acknowledgements}
  
This paper is based on observations obtained with \xmm, an ESA
science mission with instruments and contributions directly funded
by ESA Member States and the USA (NASA).

\end{acknowledgements}

\bibliographystyle{aa}
\bibliography{references}

\begin{thebibliography}{35}
\expandafter\ifx\csname natexlab\endcsname\relax\def\natexlab#1{#1}\fi

\bibitem[{{Aschwanden} \& {Parnell}(2002)}]{2002ApJ...572.1048A}
{Aschwanden}, M.~J. \& {Parnell}, C.~E. 2002, \apj, 572, 1048

\bibitem[{{Ayres}(2009)}]{2009ApJ...696.1931A}
{Ayres}, T.~R. 2009, \apj, 696, 1931

\bibitem[{{Ayres}(2014)}]{2014AJ....147...59A}
{Ayres}, T.~R. 2014, \aj, 147, 59

\bibitem[{{Baliunas} {et~al.}(1995){Baliunas}, {Donahue}, {Soon}, {Horne},
  {Frazer}, {Woodard-Eklund}, {Bradford}, {Rao}, {Wilson}, {Zhang}, {Bennett},
  {Briggs}, {Carroll}, {Duncan}, {Figueroa}, {Lanning}, {Misch}, {Mueller},
  {Noyes}, {Poppe}, {Porter}, {Robinson}, {Russell}, {Shelton}, {Soyumer},
  {Vaughan}, \& {Whitney}}]{1995ApJ...438..269B}
{Baliunas}, S.~L., {Donahue}, R.~A., {Soon}, W.~H., {et~al.} 1995, \apj, 438,
  269

\bibitem[{{Butler} {et~al.}(1986){Butler}, {Rodono}, {Foing}, \&
  {Haisch}}]{1986Natur.321..679B}
{Butler}, C.~J., {Rodono}, M., {Foing}, B.~H., \& {Haisch}, B.~M. 1986, \nat,
  321, 679

\bibitem[{{Drake} {et~al.}(2000){Drake}, {Peres}, {Orlando}, {Laming}, \&
  {Maggio}}]{2000ApJ...545.1074D}
{Drake}, J.~J., {Peres}, G., {Orlando}, S., {Laming}, J.~M., \& {Maggio}, A.
  2000, \apj, 545, 1074

\bibitem[{{Favata} {et~al.}(2004){Favata}, {Micela}, {Baliunas}, {Schmitt},
  {G{\"u}del}, {Harnden}, {Sciortino}, \& {Stern}}]{2004A&A...418L..13F}
{Favata}, F., {Micela}, G., {Baliunas}, S.~L., {et~al.} 2004, \aap, 418, L13 (Paper I)

\bibitem[{{Favata} {et~al.}(2008){Favata}, {Micela}, {Orlando}, {Schmitt},
  {Sciortino}, \& {Hall}}]{2008A&A...490.1121F}
{Favata}, F., {Micela}, G., {Orlando}, S., {et~al.} 2008, \aap, 490, 1121 (Paper II)

\bibitem[{{Feldman} {et~al.}(1997){Feldman}, {Doschek}, \&
  {Klimchuk}}]{1997ApJ...474..511F}
{Feldman}, U., {Doschek}, G.~A., \& {Klimchuk}, J.~A. 1997, \apj, 474, 511

\bibitem[{{Hara}(1996)}]{1996PhDT........62H}
{Hara}, H. 1996, PhD thesis, PhD thesis.~Natl.~Astronom.~Obs., Japan.164 pp.~,
  (1996)

\bibitem[{{Hempelmann} {et~al.}(2006){Hempelmann}, {Robrade}, {Schmitt},
  {Favata}, {Baliunas}, \& {Hall}}]{2006A&A...460..261H}
{Hempelmann}, A., {Robrade}, J., {Schmitt}, J.~H.~M.~M., {et~al.} 2006, \aap,
  460, 261

\bibitem[{{Hempelmann} {et~al.}(2003){Hempelmann}, {Schmitt}, {Baliunas}, \&
  {Donahue}}]{2003A&A...406L..39H}
{Hempelmann}, A., {Schmitt}, J.~H.~M.~M., {Baliunas}, S.~L., \& {Donahue},
  R.~A. 2003, \aap, 406, L39

\bibitem[{{Hempelmann} {et~al.}(1996){Hempelmann}, {Schmitt}, \& {St{\c
  e}pie{\'n}}}]{1996A&A...305..284H}
{Hempelmann}, A., {Schmitt}, J.~H.~M.~M., \& {St{\c e}pie{\'n}}, K. 1996, \aap,
  305, 284

\bibitem[{{Judge} {et~al.}(2003){Judge}, {Solomon}, \&
  {Ayres}}]{2003ApJ...593..534J}
{Judge}, P.~G., {Solomon}, S.~C., \& {Ayres}, T.~R. 2003, \apj, 593, 534

\bibitem[{{Mittag} {et~al.}(2016){Mittag}, {Schr{\"o}der}, {Hempelmann},
  {Gonz{\'a}lez-P{\'e}rez}, \& {Schmitt}}]{2016A&A...591A..89M}
{Mittag}, M., {Schr{\"o}der}, K.-P., {Hempelmann}, A.,
  {Gonz{\'a}lez-P{\'e}rez}, J.~N., \& {Schmitt}, J.~H.~M.~M. 2016, \aap, 591,
  A89

\bibitem[{{Orlando} {et~al.}(2000){Orlando}, {Peres}, \& {Reale}}]{opr00}
{Orlando}, S., {Peres}, G., \& {Reale}, F. 2000, \apj, 528, 524

\bibitem[{{Orlando} {et~al.}(2001){Orlando}, {Peres}, \& {Reale}}]{opr01}
{Orlando}, S., {Peres}, G., \& {Reale}, F. 2001, \apj, 560, 499

\bibitem[{{Orlando} {et~al.}(2004){Orlando}, {Peres}, \& {Reale}}]{opr04}
{Orlando}, S., {Peres}, G., \& {Reale}, F. 2004, \aap, 424, 677

\bibitem[{{Peres} {et~al.}(2000){Peres}, {Orlando}, {Reale}, {Rosner}, \&
  {Hudson}}]{porrh00}
{Peres}, G., {Orlando}, S., {Reale}, F., {Rosner}, R., \& {Hudson}, H. 2000,
  \apj, 528, 537

\bibitem[{{Pevtsov} {et~al.}(2003){Pevtsov}, {Fisher}, {Acton}, {Longcope},
  {Johns-Krull}, {Kankelborg}, \& {Metcalf}}]{2003ApJ...598.1387P}
{Pevtsov}, A.~A., {Fisher}, G.~H., {Acton}, L.~W., {et~al.} 2003, \apj, 598,
  1387

\bibitem[{{Pourbaix}(2000)}]{2000A&AS..145..215P}
{Pourbaix}, D. 2000, \aaps, 145, 215

\bibitem[{{Reale} {et~al.}(2001){Reale}, {Peres}, \&
  {Orlando}}]{2001ApJ...557..906R}
{Reale}, F., {Peres}, G., \& {Orlando}, S. 2001, \apj, 557, 906

\bibitem[{{Robrade} \& {Schmitt}(2016)}]{2016arXiv161206570R}
{Robrade}, J. \& {Schmitt}, J.~H.~M.~M. 2016, ArXiv:1612.06570

\bibitem[{{Robrade} {et~al.}(2005){Robrade}, {Schmitt}, \&
  {Favata}}]{2005A&A...442..315R}
{Robrade}, J., {Schmitt}, J.~H.~M.~M., \& {Favata}, F. 2005, \aap, 442, 315

\bibitem[{{Robrade} {et~al.}(2012){Robrade}, {Schmitt}, \&
  {Favata}}]{2012A&A...543A..84R}
{Robrade}, J., {Schmitt}, J.~H.~M.~M., \& {Favata}, F. 2012, \aap, 543, A84

\bibitem[{{Sanz-Forcada} {et~al.}(2013){Sanz-Forcada}, {Stelzer}, \&
  {Metcalfe}}]{2013A&A...553L...6S}
{Sanz-Forcada}, J., {Stelzer}, B., \& {Metcalfe}, T.~S. 2013, \aap, 553, L6

\bibitem[{{Schmitt} \& {Liefke}(2004)}]{2004A&A...417..651S}
{Schmitt}, J.~H.~M.~M. \& {Liefke}, C. 2004, \aap, 417, 651

\bibitem[{{Schmitt} {et~al.}(2014){Schmitt}, {Schr{\"o}der}, {Rauw},
  {Hempelmann}, {Mittag}, {Gonz{\'a}lez-P{\'e}rez}, {Czesla}, {Wolter}, {Jack},
  {Eenens}, \& {Trinidad}}]{2014AN....335..787S}
{Schmitt}, J.~H.~M.~M., {Schr{\"o}der}, K.-P., {Rauw}, G., {et~al.} 2014,
  Astronomische Nachrichten, 335, 787

\bibitem[{{Soderblom}(1982)}]{1982ApJ...263..239S}
{Soderblom}, D.~R. 1982, \apj, 263, 239

\bibitem[{{Stern}(1998)}]{1998ASPC..154..223S}
{Stern}, R.~A. 1998, in Astronomical Society of the Pacific Conference Series,
  Vol. 154, Cool Stars, Stellar Systems, and the Sun, ed. R.~A. {Donahue} \&
  J.~A. {Bookbinder}, 223

\bibitem[{{Tokovinin} {et~al.}(2015){Tokovinin}, {Mason}, {Hartkopf}, {Mendez},
  \& {Horch}}]{2015AJ....150...50T}
{Tokovinin}, A., {Mason}, B.~D., {Hartkopf}, W.~I., {Mendez}, R.~A., \&
  {Horch}, E.~P. 2015, \aj, 150, 50

\bibitem[{{Tsuneta} {et~al.}(1991){Tsuneta}, {Acton}, {Bruner}, {Lemen},
  {Brown}, {Caravalho}, {Catura}, {Freeland}, {Jurcevich}, {Morrison},
  {Ogawara}, {Hirayama}, \& {Owens}}]{1991SoPh..136...37T}
{Tsuneta}, S., {Acton}, L., {Bruner}, M., {et~al.} 1991, \solphys, 136, 37

\bibitem[{{Wang} \& {Dai}(2013)}]{2013NatPh...9..465W}
{Wang}, F.~Y. \& {Dai}, Z.~G. 2013, Nature Physics, 9, 465

\bibitem[{{Wargelin} {et~al.}(2017){Wargelin}, {Saar}, {Pojma{\'n}ski},
  {Drake}, \& {Kashyap}}]{2017MNRAS.464.3281W}
{Wargelin}, B.~J., {Saar}, S.~H., {Pojma{\'n}ski}, G., {Drake}, J.~J., \&
  {Kashyap}, V.~L. 2017, \mnras, 464, 3281

\bibitem[{{White} {et~al.}(1992){White}, {Skumanich}, {Lean}, {Livingston}, \&
  {Keil}}]{1992PASP..104.1139W}
{White}, O.~R., {Skumanich}, A., {Lean}, J., {Livingston}, W.~C., \& {Keil},
  S.~L. 1992, \pasp, 104, 1139

\end{thebibliography}

\end{document}